\documentclass[a4paper]{spie}  
\usepackage[utf8]{inputenc}
\usepackage[]{graphicx}
\usepackage{amssymb,amsmath,psfrag,url}
\usepackage{verbatim}
\usepackage{todonotes}
\usepackage{textcomp}

\usepackage[scientific-notation=false]{siunitx}
\newcommand{\numc}[1]{\num[exponent-product = \cdot]{#1}}
\newcommand{\curl}{\nabla \times}
\newcommand{\bbr}{\mathbb{R}}

\title{Quantifying parameter uncertainties\\ in optical scatterometry\\ using Bayesian inversion}

\author{
  Martin~Hammerschmidt,\supit{\,ab} 
  Martin~Weiser,\supit{\,b}
  Xavier~Garcia~Santiago,\supit{\,ac}\\
  Lin~Zschiedrich,\supit{\,a}
  Bernd~Bodermann,\supit{\,d}
  Sven~Burger\supit{\,ab}
\skiplinehalf
\supit{a}
JCMwave GmbH\\
Bolivarallee~22, 
D\,--\,14\,050 Berlin,
Germany
\smallskip\\
\supit{b}
Zuse Institute Berlin (ZIB)\\
Takustra{\ss}e~7,
D\,--\,14\,195 Berlin,
Germany
\smallskip\\
\supit{c}
Institut für Theoretische Festkörperphysik (KIT)\\
Wolfgang-Gaede-Stra{\ss}e 1,
D\,--\,76\,131 Karlsruhe,
Germany
\smallskip\\
\supit{d}
Physikalisch-Technische Bundesanstalt (PTB)\\
Bundesallee 100, 
D\,--\,38\,116 Braunschweig, 
Germany
\authorinfo{
Corresponding author: M.~Hammerschmidt\\
URL: http://www.jcmwave.com\\
URL: http://www.zib.de
}}
\begin{document}
\maketitle
\noindent
This paper will be published in Proc.~SPIE Vol.~{\bf 10330}
(2017) 1033004 ({\it Modeling Aspects in Optical Metrology VI}, DOI: 10.1117/12.2270596)
and is made available 
as an electronic preprint with permission of SPIE. 
One print or electronic copy may be made for personal use only. 
Systematic or multiple reproduction, distribution to multiple 
locations via electronic or other means, duplication of any 
material in this paper for a fee or for commercial purposes, 
or modification of the content of the paper are prohibited.

\begin{abstract}
We present a Newton-like method to solve inverse problems and to quantify parameter uncertainties.
We apply the method to parameter reconstruction in optical scatterometry, where we take into account a priori information and measurement uncertainties using a Bayesian approach. 
Further, we discuss the influence of numerical accuracy on the reconstruction result. 
\end{abstract}

\section{Introduction}
\label{section_introduction}
Optical scatterometry is a method to measure the size and shape of periodic micro- or nanostructures
on surfaces~\cite{Pang2012aot}.
For this purpose the geometry parameters of the structures are obtained by reproducing experimental measurement results
through numerical simulations.
Such simulations are typically performed using parameterized models and nonlinear optimization algorithms to find parameter
settings which minimize differences between the measurements and the numerically obtained result.
Often also knowledge about experimental measurement errors and model uncertainties is available, as well as prior knowledge
obtained from additional measurements. This can be used to also quantify uncertainties of the reconstructed parameter values~\cite{Gross2009euv1,Henn2012oe,Klauenberg2015m}.

Here we demonstrate an efficent method for parameter reconstruction and uncertainty quantification using a Newton method to solve the inverse problem, an
efficient finite-element based solver for the forward-problem, and a Bayesian approach for relating measurement uncertainties and prior knowledge to the reconstruction results. 
The paper is structured as follows: 
An optical scatterometry setup which serves as application example is presented in Section~\ref{section_setup}.
The reconstruction method is discussed in Section~\ref{section_reconstruction}.
Results are presented in Section~\ref{section_results}.
Limits and caveats of using difference quotients to compute partial derivatives are addressed in the Appendix. 


\section{Scatterometric measurement configuration}
\label{section_setup}
\begin{figure}[bht]
\begin{center}
\psfrag{in}{\sffamily in}
\psfrag{out}{\sffamily out}
\psfrag{}{\sffamily }
\psfrag{hsi}{\sffamily $h$}
\psfrag{hox}{\sffamily $h_\textrm{ox}$}
\psfrag{theta}{\sffamily $\theta$}
\psfrag{phi}{\sffamily $\phi$}
\includegraphics[width=.35\textwidth]{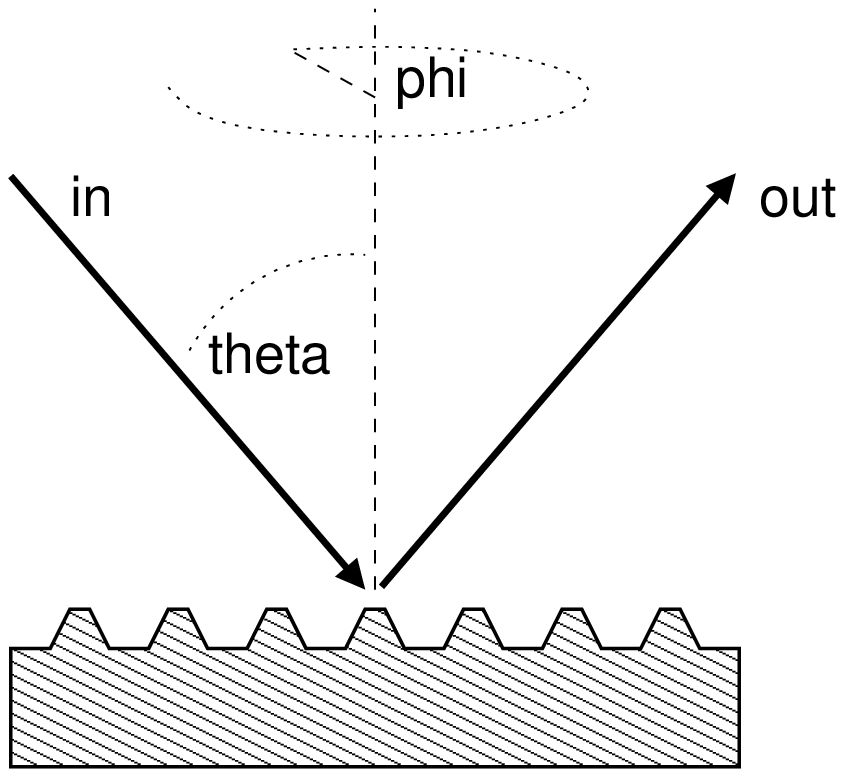} \includegraphics[width=.6\textwidth]{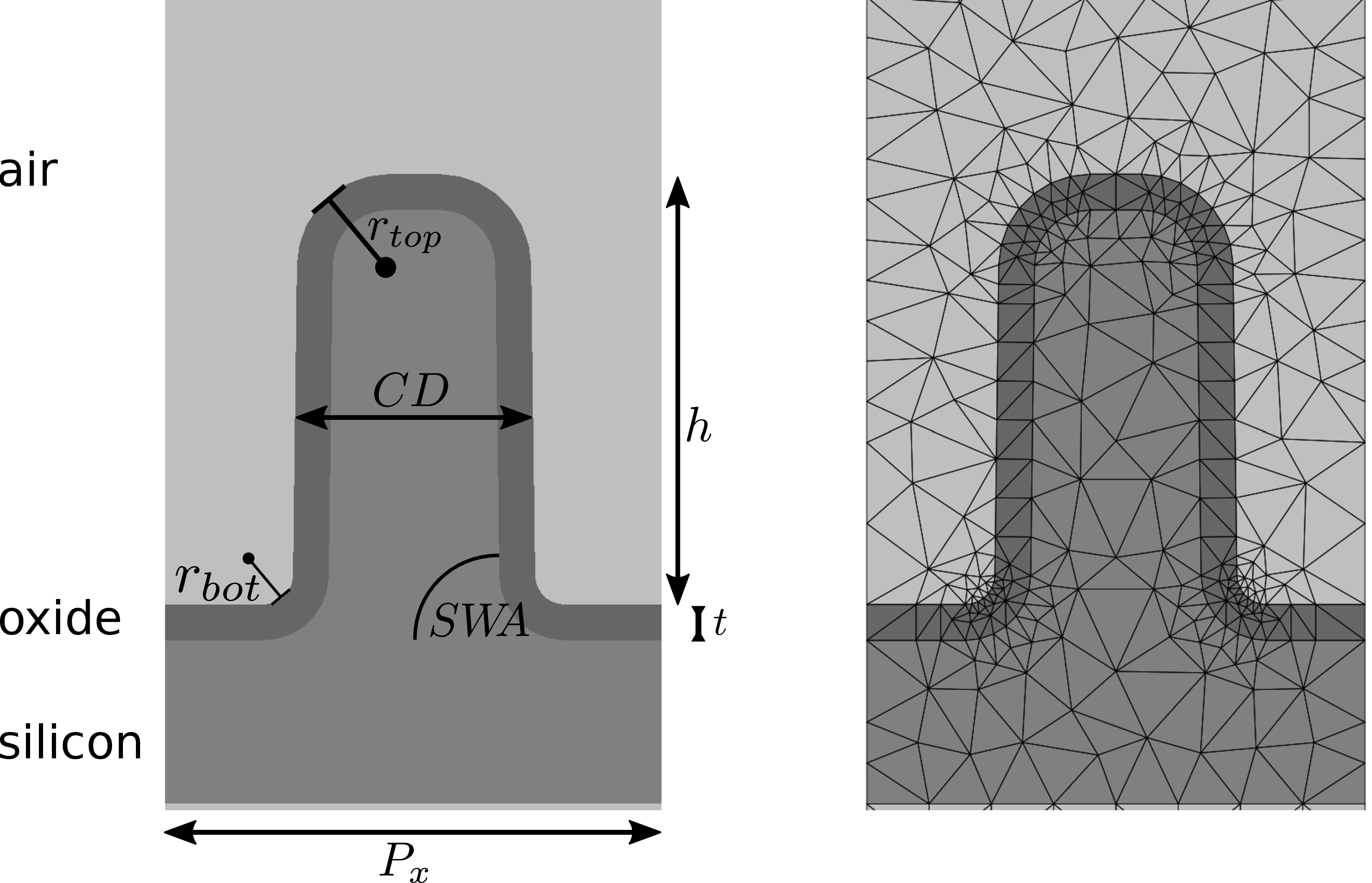}
  \caption{
{\it Left:}
Schematics of the experimental 2$\theta$ setup with incidence angle $\theta$ and azimuthal orientation $\phi$. 
{\it Center:}
Schematics of the model of a unit cell of the silicon line grating with free parameters 
line height, $h$, 
critical dimension at $h/2$, $CD$, 
oxide thickness, $h_\textrm{ox}$, 
sidewall angle $SW\!\!A$, 
top and bottom corner roundings, $r_\textrm{top}$ and $r_\textrm{bot}$.
{\it Right:}
Visualization of the triangular mesh for the FEM discretization. 
}
\label{fig_schematics_scatterometry}
\end{center}
\end{figure}

In order to test our method for solving the inverse problem we use an experimental scatterometry data set obtained at PTB.
Details of the measurement configuration have recently been reported~\cite{Wurm2011mst}.
Briefly, in the experiment, a silicon grating (1D periodic lines) with nominal pitch of $p_x=50$\,nm and nominal linewidth of $CD=25$\,nm
is used as scattering target in a goniometric setup with an inspection wavelength of $\lambda=266\,$nm. 
A schematic of the measurement is shown in Figure~\ref{fig_schematics_scatterometry} (left).
A collimated light beam at well defined polarization and angle of incidence (inclination angle $\theta$, rotation angle $\phi$) illuminates the target.
The intensity of specular reflection is recorded.
Spectra depending on inclination angle $\theta$ are recorded for S- and for P-polarization, and for rotation angles of $\phi=0$ and $\phi=90$.

The measured data set used in this study is plotted in Figure~\ref{fig:measurement_vs_sim} (circles). 

\section{Reconstruction method}
\label{section_reconstruction}
The reconstruction method described in the following employs a Bayesian perspective\cite{Sullivan2015} to compute not only the most likely geometrical parameters describing the shape of the line grating, but also to quantify the uncertainties inherent in the process. We start by describing the forward model, that is the modeling of the scattering signal given a set of parameters $x$. We address some specifics of the electromagnetic field solver and finally give a detailed insight into the Gauss-Newton method used to reconstruct the parameters.
\paragraph{Forward model.}
The forward model describes how the obtained measurements $y\in\mathbb{R}^m$ depend on the micro\-structure's parameters $x\in\mathbb{R}^n$.
The computational model aims to describe the experimental setup outlined in Section~\ref{section_setup}.
The measurement of the various combinations of incidence angles, polarizations, and azimuthal orientations are serialized and form the model evidence vector $y_M$. Each entry corresponds to a solution of the forward model for one of the given combinations. We parameterize the shape of the silicon line with a total of $n=6$ free parameters: the line height  $h$, the line width (critical dimension) at $h/2$, $CD$, the oxide layer thickness,  $h_\textrm{ox}$, the sidewall angle, $SWA$, and the top and bottom corner roundings, $r_\textrm{top}$ and $r_\textrm{bot}$. Their definitions can be found in Figure~\ref{fig_schematics_scatterometry}. In the reconstruction we allow for parameter values within large intervals describing a wide range of line shapes.
The bounds employed are listed as ranges in Table~\ref{tab:fem_scan} for all free parameters. However, not all combinations of parameters out of these intervals lead to meaningful line shapes. For example, we demand that the corner rounding radius at the top of the line is smaller than half the width at the top. This restricts the choice of parameter configurations as the parameters are no longer independent of each other.

The line grating is then modeled as a 2D unit cell in the cross section of an infinitely periodic array by employing periodic boundary conditions at the boundaries in $\pm x$ direction, with a fixed periodicity of $p_x=50$~nm. The silicon substrate and air superstrate are modeled as infinite half-spaces by employing transparent boundary conditions at the top and bottom of the computational domain shown in Figure~\ref{fig_schematics_scatterometry}.

The scattering of monochromatic light off the nanoscopic line grating is described by the linear Maxwell's equations in frequency domain.
These lead to a single second order partial differential equation
\begin{equation}\label{eq:mw}
 \curl \mu^{-1} \curl E - \omega^2 \varepsilon E = 0 
\end{equation}
where $\varepsilon$ and $\mu$ are the permittivity and permeability tensors, and $\omega$ is the time-harmonic frequency.
Here, impressed current sources are implicitely neglected.
The materials are non-magnetic ($\mu_r$=1.0) and refractive indices for silicon, the top oxide layer  and air of $n_\textrm{Si}=1.967+4.443i$, $n_\textrm{ox} = 1.7212+0.113i$ and $n_\textrm{air}=1.0$, respectively, are used.

\paragraph{Maxwell solver.} Analytical solutions of Maxwell's equations are not available on the complex parameterized geometries under consideration. Therefore, a numerical simulation of the measurement process is necessary.
We employ the finite-element (FEM) electromagnetic field solver JCMsuite~\cite{cite_jcmwave_jcmsuite,Pomplun2007pssb} 
which has been successfully used in scatterometric investigations ranging from the optical~\cite{potzick2008international} to the EUV and X-ray regimes~\cite{Scholze2007a,Soltwisch2016prb}
on 2D (e.g., line masks) and 3D (e.g., FinFETs, contact holes) scattering targets.

To solve the scattering problem outlined above, the computational domain is split into interior domain $\Omega_\textrm{CD}\subset\bbr^3$ and an exterior domain $\Omega_\textrm{ext}$
which hosts the incoming electromagnetic fields and the outgoing scattered field.
Waves incident on $\Omega_\textrm{CD}$ are added to the right hand side of Equation~\eqref{eq:mw} in the appriopriate manner\cite{Pomplun2007pssb}.
Transparent boundary conditions are realized by the perfectly matched layer (PML) method.
The finite element method employs polynomials of order $p$ with local support as ansatz functions over each element with a mesh size $h$.
Transforming Equation~\eqref{eq:mw} into a weak formulation by multiplication with a test function and subsequent integration leads to a linear equation
\begin{equation}\label{eq:fem}
 \mathcal{A}E=f
\end{equation}
for the $N$ coefficients $E$ of the FEM ansatz functions. Here, $\mathcal{A}$ is a sparse matrix  and  $f$ contains contributions of incoming fields. 

We use a workflow for the Maxwell solver starting and ending in the scripting language Matlab.
Within a typical Matlab function call,
geometrical, material and source parameters are defined,
program calls for mesh generation and solutions of the linear system, Eq.~\eqref{eq:fem}, are distributed to various computation cores of a workstation or cluster,
and the results are automatically collected and post-processed.

\paragraph{Inverse problem.} For reconstructing likely parameters and quantifying the uncertainty of the reconstruction, we take a Bayesian perspective~\cite{KaipioSomersalo2005}. In this framework, the conditional probability $\pi(x|y)$ of a certain parameter vector $x$ given the acquired measurement vector $y$, also known as posterior probability, is given by Bayes' theorem as the product of likelihood $\pi(y|x)$ and prior probability $\pi(x)$. With a sufficiently parameterized model capable of representing the actual geometry, the likelihood  describes essentially the measurement errors. Here we assume independently normally distributed errors with zero mean and diagonal covariance $\Gamma_l$:
\begin{equation}\label{eq:likelihood}
 \pi(y|x) \sim \exp\left(-\frac{1}{2} (y-y(x))^T \Gamma_l^{-1} (y-y(x))\right)
\end{equation}
The parameters defining the geometric shape of the microstructure have to stay in an admissible bounded region in order to avoid non-physical self-intersection. The admissible region can in general be defined as $X = \{ x\in\mathbb{R}^m \mid g_i(x)\ge 0, i=1,\dots,r\}$ with smooth scalar valued functions $g_i$. Imposing little bias within $X$, but assuming that parameters close to the boundary $\partial X$ are less likely to occur in practice, we define a prior density as
\begin{equation}\label{eq:prior}
 \pi(x) \sim \exp\left( \sum_{i=1}^r \mu_i \log g_i(x) \right)
\end{equation}
for penalty factors $\mu_i>0$. The larger these factors are, the more bias towards the analytic center of the feasible region is imposed by the prior density. To facilite the computation of derivatives of the prior, we rely heavily on the use of automatic differentiation\cite{Griewank2008}. 

From the resulting posterior density $\pi(x|y) = \pi(y|x)\pi(x)$ reasonable point estimates can be computed. We use the maximum posterior estimate $x_{\rm MAP} = \mathop\text{arg max}_{x\in X} \pi(x|y)$. Taking the logarithm of $\pi(x|y)$, this can be computed by solving the minimization problem
\begin{equation}\label{eq:map}
 x_{\rm MAP} = \mathop\text{arg min}\limits_{x\in X} \; F(x), \quad F(x) = \frac{1}{2}(y-y(x))^T \Gamma_l^{-1}(y-y(x)) - \sum_{i=1}^r \mu_i \log g_i(x).
\end{equation}
Here we employ a Gau\ss-Newton like method\cite{Deuflhard2006} for computing local minimizers of $F$. The objective's first and second derivatives are
\[
 F'(x) = (y-y(x))^T \Gamma_l^{-1}y'(x) - \sum_{i=1}^r  \frac{\mu_i}{g_i(x)} g'(x)
\]
and
\[
 F''(x) = y'(x)^T\Gamma_l^{-1}y'(x) + (y-y(x))^T \Gamma_l^{-1}y''(x) + \sum_{i=1}^r \left(\frac{\mu_i}{g_i(x)^2} g_i'(x)^Tg_i'(x) - \frac{\mu_i}{g_i(x)}g_i''(x)\right).
\]
For good data reproduction (i.e., small $\|y-y(x)\|$) and parameters not close to the boundary of the admissible region (i.e., small $\mu_i/g_i(x)$) the second order terms can be neglected, yielding a positive definite Hessian approximation close to the minimizer:
\[
 \hat F''(x) = y'(x)^T\Gamma_l^{-1}y'(x) + \sum_{i=1}^r \frac{\mu_i}{g_i(x)^2} g_i'(x)^Tg_i'(x)\approx F''(x)
\]
Starting at $x_0\in X$, the Gau\ss-Newton method suggests steps $\Delta x_k = -\hat F''(x_k)^{-1} F'(x_k)$ which due to positivity of $\hat F''(x)$ are descent directions, and performs updates $x^{k+1} = x_k + \alpha_k \Delta x_k$ with a step length $\alpha_k\in\mathopen]0,1]$ determined by line search such that monotone decrease of the objective is guaranteed.

The Taylor approximation
\begin{equation}\label{eq:gaussian}
F(x) \approx \hat F(x) = F(x_{\rm MAP}) + F'(x_{\rm MAP}) (x-x_{\rm MAP}) + \frac{1}{2} (x-x_{\rm MAP})^T \hat F''(x_{\rm MAP}) (x-x_{\rm MAP})
\end{equation}
provides a local Gaussian approximation of the posterior by $\pi(x|y) \approx c \exp(-\hat F(x))$ and thus allows a local uncertainty quantification in terms of the covariance $\hat\Gamma_p = \hat F''(x_{\rm MAP})^{-1}$. The corresponding variances $\hat F''(x_{\rm MAP})_{ii}^{-1}$ of the marginal posterior distributions of the individual parameters $x_i$ then give an idea of how reliably identified the parameters are. A more concise information about how well parameters are identified can be inferred from the eigenvalues and eigenvectors of the covariance $\hat \Gamma_p$, giving linear combinations of parameters that are more or less reliably estimated.


\section{Reconstruction results}
\label{section_results}
We present the results of a performed reconstruction with a random starting point in section \ref{sec:reconstruction}. In section \ref{sec:accuracy} we study the effect of numerical discretization accuracy on the reconstruction results. 

\subsection{Geometry reconstruction using a Gau{\ss}-Newton method}\label{sec:reconstruction}
We start the reconstruction algorithm outlined in the previous section at a randomly chosen starting point $x_0$ with $CD = 23.8262$\,nm, $h=43.1662$\,nm, $SW\!\!A=89.2796$\textdegree{}, $t=3.5963$\,nm, $r_\textrm{top}=9.2135$\,nm and $r_\textrm{bot}=2.9608$\,nm.
The measurement errors are assumed to be normally distributed with zero mean and relative standard deviation of 2\%, 
thus fixing $\Gamma_l$ and the likelihood in \eqref{eq:likelihood}.
The bounds employed in the prior are given as ranges in Table~\ref{tab:fem_scan} for all parameters.
The coupled constraints on the corner rounding radii are also included in the prior as outlined in Section~\ref{section_reconstruction}, Equation~\eqref{eq:prior}.

\begin{figure}[htbp]
\centering
\psfrag{phi=0, P-Pol}{\sffamily \tiny $\phi=0$, P-Pol}
\psfrag{phi=90, P-Pol}{\sffamily \tiny $\phi=90$, P-Pol}
\psfrag{phi=0, S-Pol}{\sffamily \tiny $\phi=0$, S-Pol}
\psfrag{phi=90, S-Pol}{\sffamily \tiny $\phi=90$, S-Pol}
\psfrag{I_norm}{\sffamily I\textsubscript{norm}}
\psfrag{Theta [deg]}{\sffamily $\theta$ [\textdegree{}]}
\psfrag{max}{\sffamily \tiny max}
\psfrag{# unknowns}{\sffamily \# unknowns}
\includegraphics[width=.45\textwidth]{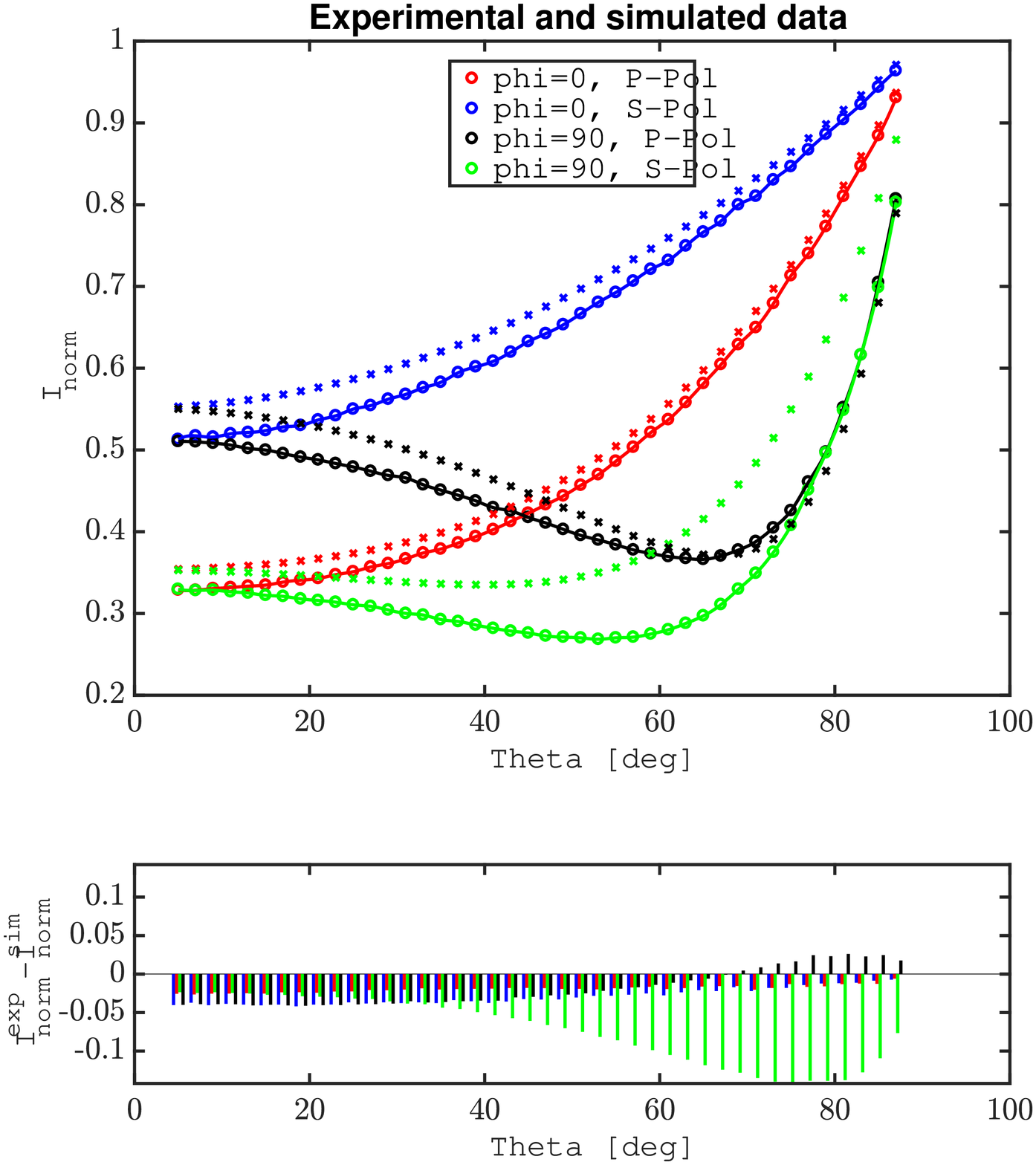}
\includegraphics[width=.45\textwidth]{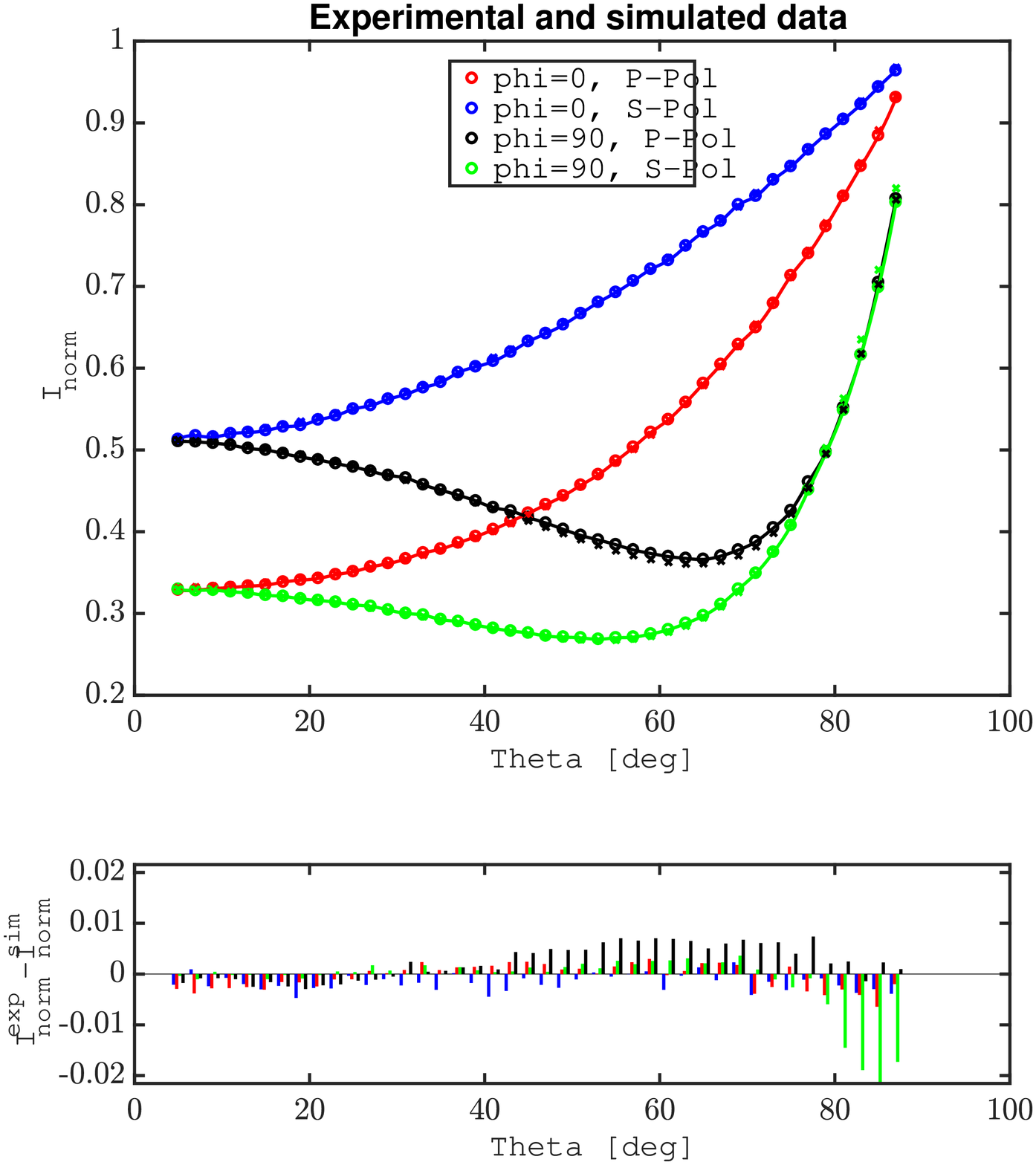}
  \caption{
{\it Left:}
Experimental data (circles and connecting lines) and simulated data for the random initial configuration $x_0$ (crosses). 
We observe a general qualitative alignment of simulated data and measured data.
The plot at the bottom shows the difference between measured and simulated signals.
Largest deviations are observed for the S-polarization, $\phi=90$\textdegree{} and large inclination angles $\theta$. 
{\it Right:}
Experimental data  (circles and connecting lines) and simulated data  for the $x_\textrm{MAP}$ configuration  (crosses). 
We observe a very good quantitative alignment of the data.
Deviations are much smaller than in the initial case.
Note the different scales of the y-axis in the deviation plots at the bottom.
}
\label{fig:measurement_vs_sim}
\end{figure}

The algorithm converges in just six Gau{\ss}-Newton iterations to the desired relative accuracy of \numc{1e-4} in the differential. 
The resulting parameter values in the maximum posterior estimate $x_\textrm{MAP}$ are
$CD = 25.3785$\,nm, $h=48.0835$\,nm, $SW\!\!A=86.9803$\textdegree{}, $t=4.9392$\,nm, $r_\textrm{top}=10.3685$\,nm and $r_\textrm{bot}= 4.7940$\,nm
(cf., Table~\ref{tab:fem_scan}). 
In Figure \ref{fig:measurement_vs_sim} the experimental and simulated intensities in the 0th diffraction order are shown as function of inclination angle~$\theta$.
The four different angular spectra refer to the different polarizations and azimuthal orientations of the illumiation.
In Fig.~\ref{fig:measurement_vs_sim} (left) the simulated data for the initial configuration $x_0$ is shown together with a plot of the differences between measured and simulated data points at the bottom.
In Fig.~\ref{fig:measurement_vs_sim} (right) the same is shown with the simulations results for $x_\textrm{MAP}$.
We observe an almost perfect alignment of the simulated data for $x_\textrm{MAP}$ and the measurements. Deviations are much smaller than in the initial case on the left.
Note the different scales of the y-axis in the deviation plots at the bottom.

From the local Gaussian approximation of the posterior distribution (cf., Equation \eqref{eq:gaussian})
we can quantify the uncertainty of the reconstructed parameter values in terms of standard deviations.
We find low standard deviations of $0.3984$\,nm and $0.1615$\,nm for the
line width $CD$ and the oxide layer thickness $t$.
The side wall angle $SW\!\!A$ has a standard deviation of $0.9988$\textdegree{}.
The height $h$ and the two corner rounding radii $r_\textrm{top}$ and $r_\textrm{bottom}$
show larger uncertainties of $2.4838$\,nm, $4.2892$\,nm  and $3.2171$\,nm respectively.
These results agree also within the uncertainty limits with a recent, different evaluation using a slightly different model~\cite{Wurm2017oe}.

\subsection{Numerical convergence study}\label{sec:accuracy}
In this section we study the impact of numerical accuracy of the forward model on the reconstructed parameter values and their uncertainties. 
The numerical accuracy of the solution of the forward problem essentially depends on the mesh width and on the polynomial degree of the ansatz functions $p$.
The small features sizes and layer thicknesses require a reasonably fine meshing of the unit cell geometry as shown in Figure~\ref{fig_schematics_scatterometry} (right).
This constitutes a lower limit on the mesh finesse provided we aim for a quality mesh which fulfills the Delaunay criterion.
Hence in this study, we keep the mesh fixed and increase the polynomial degree $p$ from 1 to 5. For each $p$ we perform the reconstruction using the same starting point $x_0$ listed above and observe the computed $x_\textrm{MAP}$. The results of this study are shown in Table~\ref{tab:fem_scan}. The results for $p>2$ seem very well converged and show only small variations in $x_\textrm{MAP}$ as well as the standard deviations compared to $p=3$. 

\begin{table}[htb]
\centering
\resizebox{\columnwidth}{!}{%
\begin{tabular}{|l|l|l|l|l|l|l|}
\hline
 & $CD$ [nm] & $h$ [nm] & $SW\!\!A$ [deg] & $t$ [nm] & $r_\textrm{top}$ [nm] & $r_\textrm{bot}$ [nm] \\
Range & (15, 35) & (40, 60)& (84, 90)& (2.5, 6.5)& (3.5, 18) & (0, 10)\\
\hline
 p=1 &   25.0365 (0.2121) & 49.6280 (1.0519) & 87.1857 (0.8990) & 4.7531 (0.1295) & 10.6567 (1.4282) & 6.9962 (1.7699)\\
 p=2 &   25.3741 (0.2840) & 48.0863 (1.8411) & 86.9918 (0.9447) & 4.9368 (0.1367) & 10.3858 (2.8578) & 4.7964 (2.3854)\\
p=3 &  25.3786 (0.4006) & 48.0828 (2.4917) & 86.9801 (1.0002) & 4.9392 (0.1621) & 10.3688 (4.3045) & 4.7927 (3.2360)\\
p=4 &  25.3785 (0.3993) & 48.0833 (2.4873) & 86.9803 (0.9994) & 4.9392 (0.1618) & 10.3687 (4.2965) & 4.7933 (3.2252)\\
p=5 &  25.3785 (0.3984) & 48.0835 (2.4838) & 86.9803 (0.9988) & 4.9392 (0.1615) & 10.3685 (4.2892) & 4.7940 (3.2171)\\
\hline
 \end{tabular}
}\caption{Reconstruction results with estimated standard deviations (in parentheses)
for numerical settings with increasing accuracy of the forward-problem ($p=1$ to $p=5$).
The admissible parameter regions in the reconstruction (Range) are indicated for all parameters.
}
\label{tab:fem_scan}
\end{table}

\begin{figure}[h]
\centering
\includegraphics[width=.6\textwidth]{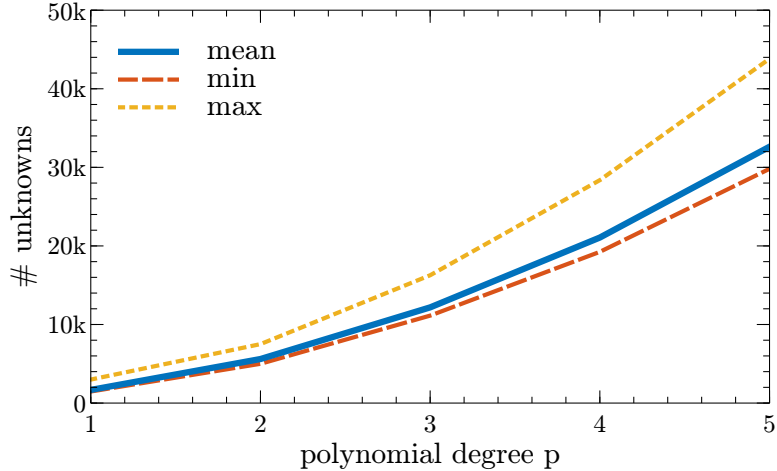}
\caption{Number of unknowns for increasing polynomial degree used in the forward-problem computations ($p=1$ to $p=5$). The mean over all simulations in the reconstruction is shown in blue, yellow and orange mark the minimum and maximum encountered. 
}
\label{fig:fem_unknowns}
\end{figure}

We distribute the independent forward problems on a parallel computing cluster.
Hence the computation time for all 168 data points considered in each iteration of the Gau{\ss}-Newton reconstruction method is limited by the time of a single FEM evaluation.
The average numerical effort required by the forward problems increases from 2047 unknowns for $p=1$ to 35863 for $p=5$.
In Figure \ref{fig:fem_unknowns} the number of unknowns is shown as a function of the polynomial degree $p$.
The mean of all 168 simulation is given by the blue line.
The PML thickness is automatically adjusted by the solver JCMsuite\cite{cite_jcmwave_jcmsuite} by considering the angle of incidence,
the material distribution in the exterior and the illumination wavelength.
In combination with a remeshing for every iteration in the reconstruction, this leads to differences in the number of degrees of freedom over the angle of incidence scan.
In Figure \ref{fig:fem_unknowns} orange and yellow lines mark the minimum and maximum the degrees of freedom observed in all simulations performed during the reconstruction.
The solution of these relatively small FEM problems including post processing takes at most 7 seconds of CPU time.
Based on these findings it seems reasonable to choose $p=3$ as a trade off between numerical accuracy and computation times of $\approx 3$ seconds of CPU time.

\begin{figure}[htp]
\centering
\includegraphics[width=\textwidth]{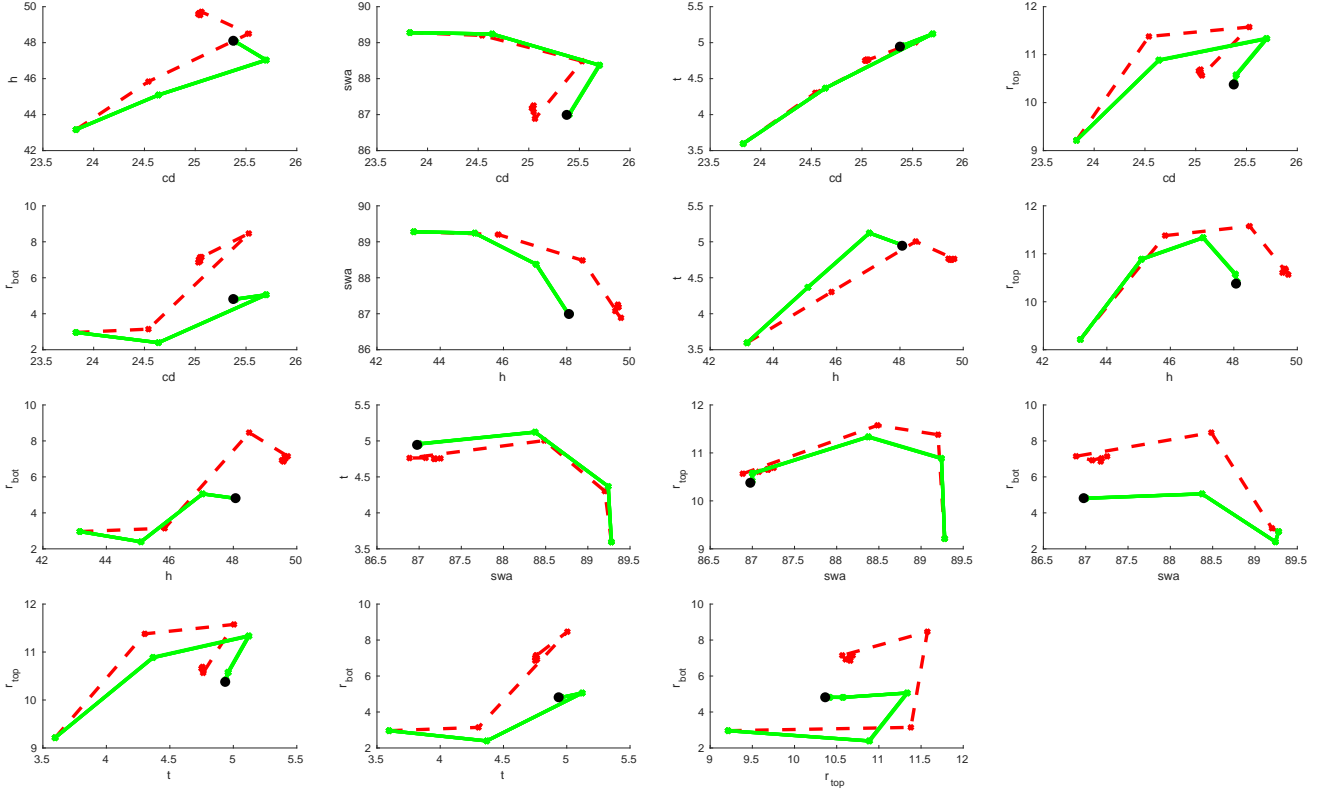}
\caption{
Pair-wise projections of the trajectories of the Gau{\ss}-Newton method in parameter space.
Converged trajectories for $p>=2$ are shown in green whereas the trajectory for $p=1$ is shown in red.
The iterates are indicated by crosses which for the green trajectories are almost identical.
The location of $x_\textrm{MAP}$ is indicated by a black dot covering the last iterates. }\label{fig:fem_trajectories}
\end{figure}
Investigating the influence of the polynomial degree in more detail, we look at the trajectories of the Newton method in parameter space. These are shown as pair-wise projections in Figure \ref{fig:fem_trajectories}. The graphs of the converged trajectories for $p>=2$ are shown in green whereas the trajectory for $p=1$ is shown in red. We observe no visible distinction between the converged results. All four trajectories include six data points  respectively, indicated by crosses in the plots. These are almost indistinguishable as all descent directions found by the Newton method are virtually identical and only differ slightly in step length. In most subplots the last two steps taken are too small to be resolved and are covered by the black circles marking $x_\textrm{MAP}^{p=5}$. The trajectory for $p=1$ (red, dotted line) is distinctly different as it requires 9 steps and converges toward a different estimate.

\section{Conclusion}
\label{section_conclusion}
The parameters describing the shape of a silicon line grating and associated uncertainties have been found by means of an Bayesian inverse problem solved with a Gau{\ss}-Newton method.
The agreement of experimental measurements and the numerical model is very good.
The Bayesian perspective and a quadratic approximation of the posterior distribution allows for a quantification of the parameter uncertainties in terms of standard deviations of a Gaussian centered at the reconstructed parameters.
The employed Gau{\ss}-Newton method to solve the inverse problem demonstrated good convergence properties and robustness. 
An in depth analysis of the numerical reconstruction method will follow in a subsequent paper.

\section*{Acknowledgments}
This project has received funding from the European Union’s Horizon 2020 research and innovation programme under the
Marie Sk{\l}odowska-Curie grant agreement No 675745 (MSCA-ITN-EID NOLOSS).
This project has received funding from the EMPIR programme co-financed by the Participating States and from the European Union’s
Horizon 2020 research and innovation programme under grant agreement number 14IND13 (PhotInd).

\bibliography{this_bibliography}
\bibliographystyle{spiebib}

\newpage\section*{Appendix}
\label{section_appendix}
\subsection*{Pitfalls in using difference quotients to compute partial derivatives}\label{section_difference_quotients}
Computation of partial derivatives of the electric field and derived quantities is required for both, the Newton method for solving the inverse problem,
and for local uncertainty quantification at $x_\textrm{MAP}$.
Here we  compare two methods of how to compute first and second order derivatives numerically, 
direct computation and computation using finite differences,
and we show the impact of insufficient computation of these on the reconstruction results. 

For direct computation the system matrix can be employed to compute partial derivatives $\frac{\partial E}{\partial x_i}$ efficiently as reported previously\cite{Burger2013al}.
This can be seen easily by differentiating the linear system $AE=f$ which gives
\begin{equation}
 \frac{\partial E}{\partial x_i} = A^{-1} \left(\frac{\partial f}{\partial x_i} - \frac{\partial A}{\partial x_i}E   \right). \label{eq:fem_diffs}
\end{equation}
Similarly, higher-order partial derivatives are accessible recursively.

Alternatively, a finite-difference scheme can be used: 
\begin{equation}
 \frac{d f}{d x_i} =  \frac{f(x+\delta x_i)- f(x-\delta x_i)}{2 \delta x_i}. \label{eq:fd}
\end{equation}
\begin{equation}
 \frac{d^2 f}{d x_i^2} =  \frac{f(x+\delta x_i) - f(x) + f(x-\delta x_i)}{(\delta x_i)^2}. \label{eq:fd2}
\end{equation}
\begin{equation}
 \frac{d^2 f}{d x_i d x_j} =  \frac{f(x+\delta x_i+\delta x_j) -f(x-\delta x_i+\delta x_j) - f(x+\delta x_i-\delta x_j) + f(x-\delta x_i-\delta x_j)}{4 \delta x_i \delta x_j}. \label{eq:fd2q}
\end{equation}

It can be shown that from a computational point of view it is generally more efficient to directly compute the partial derivatives than to compute additional data points for difference quotients.
Irregardless of their computational efficiency we want to investigate the influence of finite difference approximations to the partial derivatives.
To this extent, we have repeated the study of Section~\ref{section_results} by replacing the computation of the second order partial derivatives used to determine the standard deviations with a finite difference approximation.

\begin{table}[h]
\centering
\resizebox{\columnwidth}{!}{%
\begin{tabular}{|l|l|l|l|l|l|l|}
\hline
 & $CD$ [nm] & $h$ [nm] & $SW\!\!A$ [deg] & $t$ [nm] & $r_{\rm top}$ [nm] & $r_{\rm bot}$ [nm] \\
\hline
 p=1 &  25.0106 (-) & 49.7642 (0.8392) & 86.7925 (-) & 4.7325 (-) & 10.3226 (0.1500) & 6.4657 (1.2493)\\
 p=2 &    25.3728 (0.3429) & 48.0927 (1.2692) & 86.9852 (1.3253) & 4.9361 (0.1858) & 10.3824 (1.9281) & 4.7951 (2.2815)\\
 p=3 &    25.3788 (0.3472) & 48.0824 (1.2579) & 86.9810 (1.3285) & 4.9393 (0.1867) & 10.3694 (1.9056) & 4.7939 (2.3644)\\
 p=4 &    25.3787 (0.3481) & 48.0827 (1.2572) & 86.9811 (1.3309) & 4.9393 (0.1871) & 10.3690 (1.9111) & 4.7945 (2.3680)\\
 p=5 &    25.3787 (0.3493) & 48.0829 (1.2575) & 86.9811 (1.3344) & 4.9393 (0.1876) & 10.3689 (1.9234) & 4.7952 (2.3685) \\
\hline
 \end{tabular}
}\caption{Reconstruction results with estimated standard deviations (in brackets)
for numerical settings with increasing accuracy of the forward-problem.
The standard deviations are not reliable in this case as they are computed in a not sufficiently converged finite difference scheme. 
}
\label{tab:fem_scan_diffq}
\end{table}

The results of this scan are listed in Table \ref{tab:fem_scan_diffq}. Unsurprisingly we get the same values for $x_\textrm{MAP}$ as the quasi Newton iterations are not influenced by our change.
The standard deviations however are changing.
In case of low numerical accuracy ($p=1$) the standard deviations for the $CD$, the $SW\!\!A$ and $t$ are not reliably computed. 
Even though we seem to observe convergence in this quantity with increasing $p$, we note difference to the values found with exact second order partial derivatives listed in Table \ref{tab:fem_scan}.
The standard deviations for $CD$ and $t$ are relatively similar whereas the larger standard deviations for $h$, $r_\textrm{top}$ and $r_\textrm{bot}$ change drastically by factors of almost two. 

\begin{figure}[h]
\centering
\psfrag{delta}{\sffamily \small $\delta x_i$ [nm / \textdegree{}]}
\psfrag{YLABEL1}{\sffamily \small rel. error in $\frac{\partial I}{\partial x_i}$}
\psfrag{YLABEL2}{\sffamily \small rel. error in $\frac{\partial I}{\partial x_i}$}
\psfrag{XXXXXX1}{\sffamily \tiny $ \partial$ I / $\partial$ CD}
\psfrag{XXXXXX2}{\sffamily \tiny $ \partial$ I / $\partial$ h}
\psfrag{XXXXXX3}{\sffamily \tiny $ \partial$ I / $\partial$ SWA}
\psfrag{XXXXXX4}{\sffamily \tiny $ \partial$ I / $\partial$ t}
\psfrag{XXXXXX5}{\sffamily \tiny $ \partial$ I / $\partial$ r\textsubscript{top}}
\psfrag{XXXXXX6}{\sffamily \tiny $ \partial$ I / $\partial$ r\textsubscript{bot}}
\includegraphics[width=.95\textwidth]{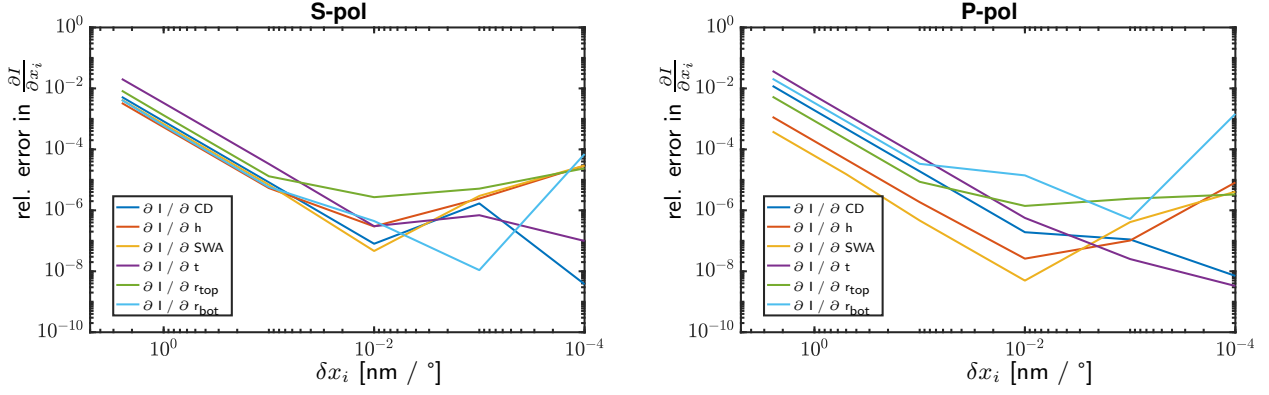}
\caption{
Convergence of first order partial derivatives with perturbation $\delta x_i$ for S- and P-polarization.
}
\label{fig:diff_quotient}
\end{figure}

{\bf Convergence of first order derivatives:}
We studied the convergence of first order partial derivatives of the 0th order intensity using equation~\eqref{eq:fd} with respect to the ones obtained with equation \eqref{eq:fem_diffs}
for one specific setting of the forward problem ($p=5$, $\theta=45$\textdegree{}).
The results are shown in Figure \ref{fig:diff_quotient}.
The convergence with respect to the perturbation $\delta x_i$ is shown for both polarizations, the different colors indicate the parameters.
For both polarizations exponential convergence can be observed for larger values of $\delta x_i$.
At a relative accuracy of approximately \numc{1e-5} the behavior changes and the relative errors remain at or sligthly below this value. 

\begin{figure}[h]
\centering
\psfrag{delta}{\sffamily \small $\delta x_i$ [nm / \textdegree{}]}
\psfrag{YLABEL1}{\sffamily \small rel. error in $\frac{\partial^2 I}{\partial x_i^2}$}
\psfrag{YLABEL2}{\sffamily \small rel. error in $\frac{\partial^2 I}{\partial x_i^2}$}
\psfrag{XXXXXX1}{\sffamily \tiny $ \partial^2$ I / $\partial$ CD\textsuperscript{2}}
\psfrag{XXXXXX2}{\sffamily \tiny $ \partial^2$ I / $\partial$ h\textsuperscript{2}}
\psfrag{XXXXXX3}{\sffamily \tiny $ \partial^2$ I / $\partial$ SWA\textsuperscript{2}}
\psfrag{XXXXXX4}{\sffamily \tiny $ \partial^2$ I / $\partial$ t\textsuperscript{2}}
\psfrag{XXXXXX5}{\sffamily \tiny $ \partial^2$ I / $\partial$ r\textsubscript{top}\textsuperscript{2}}
\psfrag{XXXXXX6}{\sffamily \tiny $ \partial^2$ I / $\partial$ r\textsubscript{bot}{2}}
\includegraphics[width=.95\textwidth]{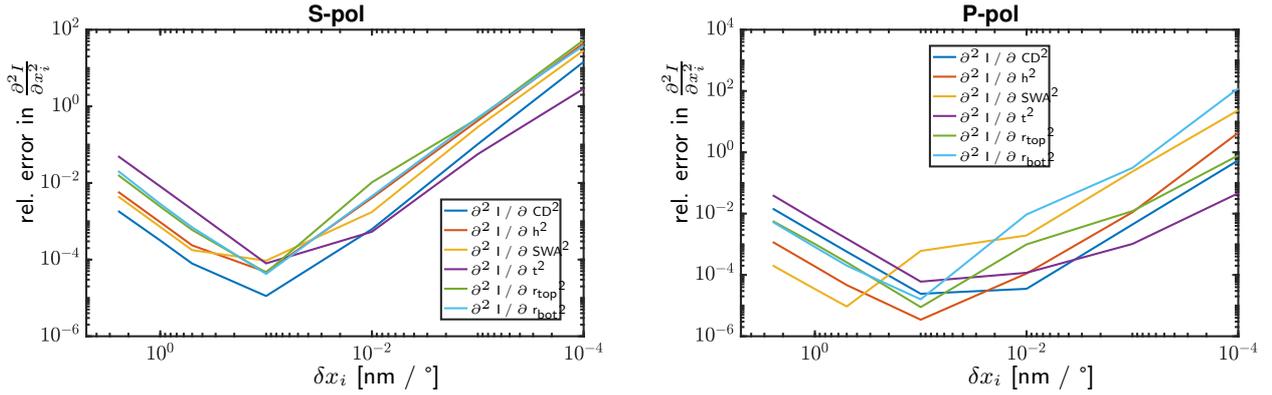}
\caption{
Convergence of second order partial derivatives with perturbation $\delta x_i$ for S- and P-polarization.
}
\label{fig:diff_quotient2}
\end{figure}

{\bf Convergence of second order derivatives:}
We studied the convergence of second order derivatives in the same setup as for the first order derivatives.
The convergence with $\delta x_i$ is shown in Figure \ref{fig:diff_quotient}.
All quantities exhibit exponential convergence provied the perturbation $\delta x_i$ is larger than \numc{1e-1}.
The relative error for this value is below \numc{1e-4} except for $\frac{d^2 I}{d SWA^2}$ where it is sligthly larger for the P-polarization.
However, instead of decreasing further the relative error increases exponentially with smaller $\delta x_i$.
This can be explained by on one hand by the smaller and smaller changes introduced in the measured signal (0th order reflected intensity)
which is captured by the numerators in equation \eqref{eq:fd2} and \eqref{eq:fd2q}.
The scaling by the denominator then leads to the exponential increase observed in Figure \ref{fig:diff_quotient2}.

These studies serve to explain the differences observed between the standard deviations in Tables \ref{tab:fem_scan_diffq} and \ref{tab:fem_scan}:
The employed perturbations ($\delta x_i \in [\numc{1e-2},\numc{4e-2}]$ depending on the magnitude of the different parameters) are large enough to provide accurate approximation of the first order partial derivatives thus ensuring the convergence of the method. However, the computational accuracy of the second order partial derivatives are insufficient for these perturbations, especially for those parameters with very small second order derivatives. The computed standard deviations in Table \ref{tab:fem_scan_diffq} thus deviate more significantly from those listed in Table \ref{tab:fem_scan}.

\end{document}